\documentclass[sn-mathphys-num]{sn-jnl}

\usepackage[utf8]{inputenc}
\usepackage{graphicx}%
\usepackage{multirow}%
\usepackage{amsmath,amssymb,amsfonts}%
\usepackage{amsthm}%
\usepackage{mathrsfs}%
\usepackage[title]{appendix}%
\usepackage{xcolor}%
\usepackage{textcomp}%
\usepackage{manyfoot}%
\usepackage{booktabs}%
\usepackage{algorithm}%
\usepackage{algorithmicx}%
\usepackage{algpseudocode}%
\usepackage{listings}%
\usepackage{lmodern}
\usepackage{anyfontsize}
\usepackage{graphicx}    
\usepackage{subcaption}  
\usepackage{float}       

\begin{document}

\title[A Machine Learning Tool to Analyse Spectroscopic Changes in High-Dimensional Data]{A Machine Learning Tool to Analyse Spectroscopic Changes in High-Dimensional Data}

\author[1]{\fnm{Alberto} \sur{Martinez-Serra}}
\email{amartinezserra@rcsi.com}
\equalcont{These authors contributed equally to this work.}

\author[2,3]{\fnm{Gionni} \sur{Marchetti}}
\email{gionnimarcheti@ub.edu}
\equalcont{These authors contributed equally to this work.}

\author[4]{\fnm{Francesco} \sur{D'Amico}} \email{francesco.damico@elettra.eu}

\author[5]{\fnm{Ivana} \sur{Fenoglio}} \email{ivana.fenoglio@unito.it}

\author[4]{\fnm{Barbara} \sur{Rossi}} \email{barbara.rossi@elettra.eu}

\author*[1]{\fnm{Marco P.} \sur{Monopoli}} \email{marcomonopoli@rcsi.com}

\author*[2,3]{\fnm{Giancarlo} \sur{Franzese}} 
\email{gfranzese@ub.edu}

\affil[1]{\orgdiv{Department of Chemistry}, \orgname{Royal
College of Surgeons in Ireland (RCSI), University of Medicine and
Health Sciences}, \orgaddress{\street{123 St Stephen's Green}, \city{Dublin}, \postcode{D02 YN77}, \country{Ireland}}}

\affil[2]{\orgname{Departament de F{\'i}sica de la Mat{\`e}ria Condensada}, \orgdiv{Universitat de Barcelona}, \orgaddress{\street{Mart{\'i} i Franqu{\`e}s 1}, \city{Barcelona}, \postcode{08028}, \country{Spain}}}

\affil[3]{\orgname{Institut de Nanoci{\`e}ncia i Nanotecnologia}, \orgdiv{Universitat de Barcelona}, \orgaddress{\street{Diagonal 645}, \city{Barcelona}, \postcode{08028}, \country{Spain}}}

\affil[4]{\orgname{Elettra Sincrotrone Trieste}, \orgaddress{\street{S.S. 114 km 163.5}, \city{Basovizza}, \postcode{34149 Trieste}, \country{Italy}}}

\affil[5]{\orgname{Department of Chemistry}, \orgdiv{University of Turin},
\orgaddress{\street{Via Pietro Giuria 7}, \postcode{10125 Turin}, \country{Italy}}}


\abstract{When nanoparticles (NPs) are introduced into a biological solution, layers of biomolecules form on their surface, creating a corona. Understanding how the protein's structure evolves into the corona is essential for evaluating the safety and toxicity of nanotechnology. However, the influence of NP properties on protein conformation is not well understood. In this study, we propose a new method that addresses this issue by analyzing multi-component spectral data (UV Resonance Raman, Circular Dichroism, and UV absorbance) using machine learning (ML). We apply the method to fibrinogen, a crucial protein in human blood plasma, at physiological concentrations while interacting with hydrophobic carbon or hydrophilic silicon dioxide NPs, revealing striking differences in the temperature dependence of the protein structure between the two cases. Our unsupervised ML method a) does not suffer from the challenges associated with the \emph{curse of dimensionality}, and b) simultaneously handles spectral data from various sources.  The method offers a quantitative analysis of protein structural changes upon adsorption. It enhances the understanding of the correlation between protein structure and NP interactions, which could support the development of nanomedical tools to treat various conditions.}

\keywords{Unsupervised Machine Learning, Manifold Reduction, Similarity Metrics, Clustering, Protein Structure, Biomolecular Corona, Spectroscopy, Hydrophobic Nanoparticles, Hydrophilic Nanoparticles}

\maketitle

\section{Introduction}\label{sec1}
 Nanoparticles (NPs) are revolutionizing various scientific fields, from medicine to environmental science, due to their unique physical and chemical properties. As these particles interact with biological systems, understanding their behavior at the molecular level becomes crucial \cite{soliman2024understanding}. There is considerable experimental evidence that the NPs' penetration of biological barriers, such as the blood-brain barrier, the intestines, and the lungs, is an active process that depends on the composition of the external biological solution and the biomolecular corona that forms as a result of the bio-nano interactions \cite{waheed2022engineering}.  This surprising dependence clearly distinguishes nanomaterials from chemical molecules \cite{monopoli2012biomolecular}. Indeed, proteins, glycans, and lipids adhere so strongly to the NP surface that the exchange times with the solution are extremely long. Consequently, the biological identity of the particles depends mainly on the bio-corona instead of the NP chemistry itself \cite{dawson2021current}.

\begin{figure}
\centering
  \includegraphics[width=\linewidth]{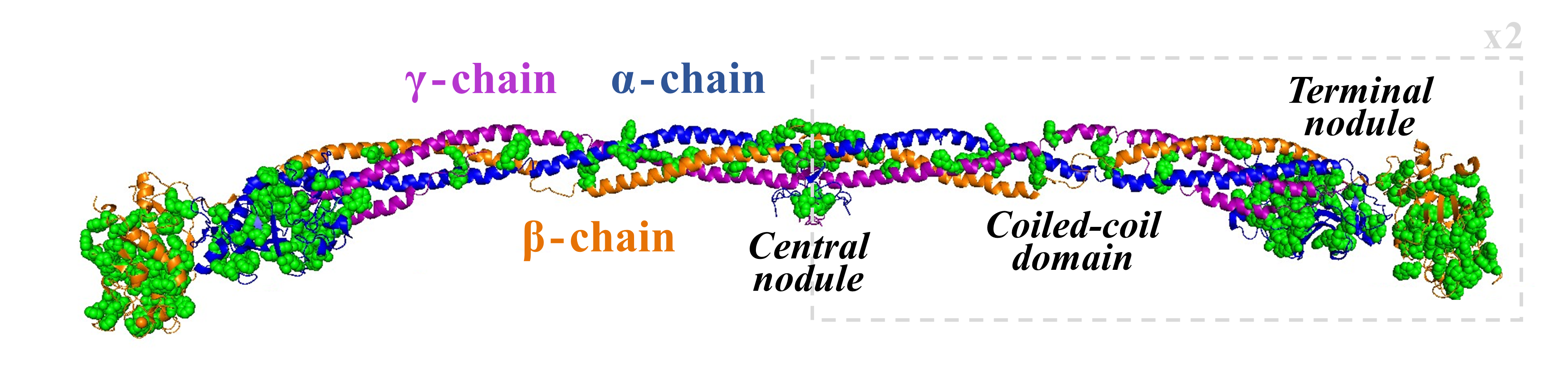}
  \vspace{1mm}
  \caption{3D structure of Fib with atomic resolution determined by X-ray crystallography (PDB Entry: 3GHG). Different colors represent the various polypeptide chains of Fib: the 
$\alpha$ (blue), 
$\beta$ (orange), and 
$\gamma$ (pink) chains. It is a symmetric protein, with chains joined together at the central node, and terminal nodes for the $
\beta$ and $\gamma$ chains. The aromatic amino acids are represented as green spheres and are mostly located in the terminal nodules and the central node, and much less in the alpha-helix. The dashed gray rectangle labeled ``x2" highlights that the protein has two symmetric parts.}
  \label{fgr:fibcartoon}
\end{figure}

Within the main constituents of the corona formed when NPs are exposed to human plasma, we find fibrinogen (Fib) 
\cite{soddu2020identification}, one of the most abundant glycoproteins in human blood plasma with a concentration of about 1.5-4 g/L.\cite{kamath2003fibrinogen}. Synthesized in the liver by hepatocytes, it is essential for the coagulation cascade and also plays a significant role in blood viscosity, blood flow, and various other biological functions \cite{herrick1999fibrinogen}. Human Fib is formed by three polypeptide chains known as 
$\alpha$, 
$\beta$, and 
$\gamma$ chains, which remain linked by disulfide bonds in the terminal domains (Fig. \ref{fgr:fibcartoon}). Their respective molecular masses are approximately 66.5, 52.0, and 46.5 kDa. The $
\beta$ and $
\gamma$ chains have two outer D domains or nodules, which connect to a central E domain through an alpha-helix segment. Along with the co- and post-translational N-linked carbohydrates added to the $
\beta$ and $
\gamma$ chains, the protein has a total molecular mass of around 340 kDa \cite{weisel2017fibrin}. The domains at the extremes of the Fib structure are essential for understanding the biological role of the molecule in the human body. They enclose several binding sites that are involved in crucial processes, such as Fib conversion to fibrin and its cross-linking \cite{mosesson2001structure}. The $
\alpha$ chain lacks these terminal nodules and essentially folds the other two chains, providing stability to the 3D structure. During the last decades, many studies have focused on the correlation of unusually elevated levels of Fib with an increased risk of experiencing health disorders, such as cardiovascular diseases \cite{kamath2003fibrinogen}. 
Furthermore, Fib has been shown to mediate adverse responses when interacting with biomaterials \cite{Thevenot:2008aa} or nanobiomaterials \cite{Ong:2015aa}. Its adsorption and conformational changes can disrupt plasma homeostasis, trigger immune responses, and promote platelet activation \cite{deng2011nanoparticle, wang2020effects}, with relevant implications for cancer nanotherapeutics and inflammatory disease contexts \cite{lim2024overcoming}.

The evolution of the protein corona has been described by theory \cite{vilaseca2013understanding} and confirmed experimentally \cite{xie2018revealing, vilanova2025characterizing}, and its dynamic changes are crucial for nanotoxicology and nanomedicine \cite{vilanova2016understanding}. However, the way the NP surface influences the protein structural changes within the corona remains debated \cite{McLoughlin19396, C5EN00054H, Kharazian:2018aa, Guo:2024aa}. For instance, theory predicts that proteins near a hydrophobic interface should unfold and aggregate less than those in bulk, and that increasing temperature does not facilitate unfolding and reduces aggregation \cite{march2021protein}. Conversely, on a hydrophilic NP, the prediction is that proteins adsorb while remaining folded, and as temperature increases, they unfold but do not desorb \cite{Fauli:2023aa}. 

Recent experiments have confirmed that the surface chemistry of NP strongly influences corona formation and protein response. Hydrophobic surfaces promote hemolysis and platelet adhesion \cite{saha2014protein, soddu2020identification}, while hydrophilic surfaces allow for more tunable coronas \cite{martinez2025multiple}. Experiments using NPs with different ligand chemistries \cite{dridi2024probing, fernandez2025functional} and various sizes \cite{sakaguchi2025changes} further show how polarity and curvature affect protein adsorption and secondary structure.

Machine learning (ML) has emerged as a powerful tool for uncovering patterns in complex biological datasets, particularly where subtle spectral differences are difficult to detect visually. For instance, recent research demonstrates that ML models can accurately classify different stages of anesthesia based on near-infrared spectroscopy data \cite{liu2024classification}. 

Here, we combine experimental techniques with machine learning (ML) analysis to investigate how Fib undergoes structural changes across various temperatures when exposed to carbon (CNP) and silicon dioxide (SiO$\mathrm{_2}$) NPs. We focus on NPs of similar size but with different surface chemistries to isolate the effects of hydrophobicity and hydrophilicity on protein behavior. 

To probe protein structural changes, we employ a combination of spectroscopic techniques: UltraViolet Resonance Raman (UVRR) spectroscopy \cite{Asher1993},  
Circular Dichroism (CD) \cite{Fasman1996}, and UV absorption spectroscopy \cite{pignataro2020}. 
Although UVRR, CD, and UV absorption provide highly specific spectral fingerprints, differences between conditions can be subtle and not easily visible to the naked eye. Traditional workflows that depend on visual comparison and manual peak fitting or deconvolution are operator-dependent and can yield inconsistent results, especially when spectra are noisy or have overlapping features. Furthermore, modern instruments regularly generate large, multi-modal datasets, making thorough manual analysis impractical. These limitations motivate our unsupervised ML approach, which enables us to extract quantitative, reproducible descriptors of conformational change from complex, multi-modal datasets, while minimizing operator bias and the constraints of manual spectral analysis. In fact, we chose unsupervised ML because there are no reliable labels for conformational states of our experimental system in these spectra, and supervised models would risk bias.

We examine high-dimensional, multi-component spectroscopic data using a novel unsupervised ML workflow. This approach is designed to quantify and visualize protein conformational changes as temperature varies. Spectroscopic datasets often contain subtle structural transitions embedded within complex, high-dimensional data that are affected by noise and collinearity. These data typically include hundreds to thousands of correlated variables, characterized by information redundancy, small sample sizes, and significant uncertainty, resulting in the \emph{curse of dimensionality}, which hinders effective analysis and interpretation \cite{Song:2024aa}.

Noise arising from instrumental limitations, environmental fluctuations, and sample variability can distort the data's intrinsic structure, reducing the effectiveness of traditional dimensionality reduction techniques \cite{Zhao:2023aa}. Additionally, collinearity among spectral bands inflates the feature space without adding meaningful information, increasing the risk of overfitting and limiting model generalizability \cite{10.3389/fpls.2023.1260772}.
To address the computational challenges of integrating multispectral data, as described above, advanced ML methods are essential for uncovering the underlying low-dimensional structure of the data. Manifold learning techniques like t-distributed stochastic neighbor embedding (t-SNE) have proven to be powerful tools, providing nonlinear dimensionality reduction capabilities that preserve both local and global data relationships, along with computational efficiency \cite{Yi:2024aa}.
The novelty of our proposed ML model rests on two key algorithms: the noisy-resistant Wasserstein distance \cite{villani2008, peyre2020} and t-SNE \cite{hinton2002, vandermaaten08, Kobak2019}. 

Our analysis reveals distinct effects of NP surface chemistry on protein structural disorder and aggregation. Fib adsorbed onto CNPs maintains a partially unfolded structure that is only weakly dependent on temperature, while Fib on SiO$_2$NPs exhibits a temperature-dependent increase in disorder similar to free Fib, with a marked transition around 50$^\circ$C. These findings, supported by unsupervised ML analysis, provide experimental confirmation of theoretical predictions and underscore the importance of NP surface properties in protein adsorption and stability.


\section{Materials and Methods}\label{secA1}
\subsection{Materials}
We use monodisperse nanoparticles of similar size but differing surface chemistry. 
SiO$_2$NPs of 0.1 \textmu m (stock concentration of 50 mg/mL) were provided by Kisker Biotech GmbH (Germany). CNPs were synthesized in the lab. 
CNPs and SiO$_2$NP have a comparable geometrical and hydrodynamic diameter to prevent any effects caused by a differing surface curvature. 

Both exhibit a negative zeta potential at neutral pH \cite{soddu2020identification, Fenoglio:2024aa}, indicating the presence of weakly acidic groups on the surface. 
In the case of CNPs, these species include acidic carboxylic or phenolic groups formed during synthesis, while for SiO$_2$NP, they consist of surface hydroxyl groups (silanols, Si-OH). 

The density of acidic groups on the NP surfaces is expected to be in the same range for both samples (around 4-6 groups per nm$^2$) \cite{soddu2020identification}. However, the two surfaces exhibit significant differences. While in CNPs the oxygenated moieties are bound to a hydrophobic surface composed of C-C bonds, in the case of silica, silanols are attached to a more hydrophilic framework made up of polar Si-O-Si bridges. Therefore, while SiO$_2$NPs are hydrophilic, CNPs are more hydrophobic.

Phosphate buffer saline (PBS) tablets, Eppendorf LoBind microcentrifuge tubes, Fibrinogen from human plasma (F3879), and D-(+)-Sucrose (99.9\%) were purchased from Sigma-Aldrich (Ireland). One PBS tablet was dissolved in 200 mL of ultrapure water to obtain a 0.01 M phosphate buffer, 0.0027 M potassium chloride, and 0.137 M sodium chloride solution (pH 7.4 at 25$^\circ$C). TEM grids of Formvar / Silicon monoxide 200 mesh with copper approximate grid hole size of 97 $\mathrm{\mu}$m, were purchased from Ted Pella (USA).

\subsection{Methods}\label{secA2}

\subsubsection{Experimental details}
\textit{\textbf{Carbon nanoparticle synthesis}} \newline
CNPs of 100 nm nominal diameter were synthesised using the method developed by Kokalari et al. \cite{kokalari2019pro}. They were produced by the carbonization of glucose in water, consisting mainly of amorphous carbon with graphitic residues incorporated in the amorphous matrix. Compared to other carbon nanomaterials, they have the advantage of being produced with precise control of the dimensions \cite{D4NA00923A}.
Moreover, they exhibit a negative charge on the surface at neutral pH, due to the presence of acidic carboxylic or phenolic groups exposed on the surface. These groups allow one to tune the hydrophobicity of the NPs.

\medskip
\textit{\textbf{Nanoparticle characterization}}\newline
Dynamic Light Scattering (DLS) measurements were performed using Zetasizer Nano ZS (Malvern). The sample cuvettes were equilibrated at 25$^\circ$C for 90 s. For each measurement, the number of runs and duration were automatically determined and repeated three times. Data analysis was performed according to standard procedures and interpreted through a cumulant expansion of the field autocorrelation function to the second order (Fig. S1 in the Electronic Supporting Information, or ESI).

Differential Centrifugal Sedimentation (DCS) experiments were performed with a CPS Disc Centrifuge DC24000, using the standard sucrose gradient 8-24\% (Analytik Ltd.). The densities were set as 1.4 and 2 g/$\mathrm{cm^3}$ for CNP and SiO$_2$NP, respectively. A 544nm PVC calibration standard was used for each sample measurement. The time taken for spherical particles with homogeneous density to travel from the center of the disk to the detector can be directly related to the apparent particle size, assuming spherical shape. 

Transmission electron microscopy (TEM) images were obtained with a JEOL JEM-1400PLUS transmission electron microscope operating at an acceleration voltage of 120 kV. The statistical analysis was carried out using ImageJ, and the average size and standard deviation were measured on 103 NPs.

\medskip
\textit{\textbf{Protein Corona obtention}}\newline
CNPs from a 4 mg/mL stock solution and SiO$_2$NPs from a 50 mg/mL stock solution were introduced into low protein-binding 1.5 mL microtubes and diluted to 2 mg/mL, respectively. We weighed 4 mg of Fib using a microbalance and diluted it in PBS to prepare the protein stock solution, which was later transferred into the NP solution to achieve a diluted concentration of 2 mg/mL. Consequently, NP-protein corona complexes formed after 10 minutes of incubation.

A total volume of $500\mathrm{\mu}$L from the final solution was transferred to a 1 mm cell to perform the UVRR. For the DCS, $100\mathrm {\mu}$L from the final sample was used at each run. For the DLS (Fig. S2 in ESI), the in-situ corona sample was diluted ten times to get an optimal attenuation. For the CD and UV absorption measurements, the sample was diluted four times to prevent signal saturation.

\subsubsection{UV Resonant Raman measurements}

UVRR spectra were collected by using the synchrotron-based UVRR set-up available at the BL10.2-IUVS beamline of Elettra Sincrotrone Trieste (Italy) \cite{d2013uv}. The excitation wavelength at 226 nm was obtained by setting the energy of the emission of the Synchrotron Radiation (SR) light by regulating the undulator gap aperture. The incoming SR light was further monochromatized using a Czerny-Turner monochromator (750 nm focal length, Acton SP2750, Princeton Instruments, Acton, MA, USA), equipped with a holographic grating featuring 3600 grooves/mm. The Raman signal of the protein solutions was collected in back-scattered geometry and analyzed using a single-pass Czerny-Turner spectrometer (Trivista 557, Princeton Instruments, with a focal length of 750 mm) equipped with a holographic grating at 1800 grooves/mm. A thermoelectrically cooled CCD camera optimized for the UV range was used to detect the signal at the spectrometer's output. The resolution was set at about $1.6\mathrm{cm^{-1}}$/pixel. Cyclohexane (spectroscopic grade, Sigma-Aldrich) was used for calibrating the wavenumber scale. The final radiation power on the samples was about $20\mathrm{\mu}$W. To avoid photodegradation during the measurements, the samples were continuously oscillated with a 1 Hz frequency and a path length of 1 mm. For the collection of temperature-dependent measurements (in the temperature range of 22-86$^\circ$C with a step of 4$^\circ$C) on protein solutions, a sample holder equipped with a thermal bath coupled to a resistive heating system was used to control the temperature with a stability of $\pm 0.1^\circ$C. The solution of protein with and without NP was measured at a concentration of 2 mg/mL (the same weight/molar concentration was kept for protein and NP). Five individual frames were collected from each sample and averaged in order to improve the signal-to-noise ratio of each Raman spectrum. We checked that no inconsistencies, irreproducibility, or photodamaging effects were observed during data acquisition. The UVRR spectra were subtracted from an almost flat background (2nd-order polynomial baseline) and smoothed using a Savitzky-Golay smoothing filter with a 5-point window width. No normalization has been used in the representation of the experimental Raman profiles. 

\subsubsection{Circular dichroism and UV absorption measurements}

CD and UV absorption spectra were recorded using a Jasco J-810 polarimeter equipped with a plug-n-play single-cell Peltier with temperature control. We measured the protein solution samples, both with and without NP, in PBS buffer (pH 7.4) at a concentration of 0.25 mg/mL. The spectra were collected in the temperature range of 22-87$^\circ$C with a step of 2$^\circ$C. The same weight/mol concentration of NP and protein was kept. All the solutions were freshly prepared in a rectangular quartz cell with a 1 mm path length. Each CD and UV spectrum was collected in the range from 190 to 260 nm with an increment of 1 nm, a scan rate of 20 nm/min, and a 1 nm bandpass. Measurements were performed under a constant nitrogen flow, and the spectra were averaged from 4 scans. For each set of temperature-dependent measurements, the spectra of the solvent (PBS and NP solution in PBS at a concentration of 0.25 mg/mL) were subtracted from the corresponding sample spectra. The experimental UV absorption and CD profiles have been smoothed using a Savitzky-Golay smoothing filter with a 15-point window width. No normalization has been used in the representation of the experimental CD and UV absorption spectra.

\subsubsection{Data Curation}\label{Data_Curation}

The data obtained through different spectral techniques form a dataset containing UVRR, CD, and UV absorption spectra of Fib alone or in conjunction with CNP or SiO$_2$NP for each temperature T, ranging from 22 to 86$^\circ$C, taken at temperature intervals of $\Delta T=4^\circ$C. This results in an overall number of 17 spectra for each type of spectroscopic method. The UVRR spectra encompass silent regions where no useful information is available. Through spectral truncation~\cite{guo2021}, these regions are removed, retaining the following five intervals (see Section~\ref{UVRR}), which are reported in Table~\ref{tab:table1}.

\begin{table}
\caption{The  UVRR spectral intervals in cm$^{-1}$ unit under scrutiny together with the corresponding significant feature of interest and number of spectral components (or spectral intensities) $n_c$. Note that the interval  $1660-1670$ cm$^{-1}$ corresponds to a region of bulk Fib where a peak disappears altogether.}
\label{tab:table1}
\centering
\begin{tabular}{c @{\hskip 1in} c @{\hskip 1in} c} 
\toprule
 Interval (cm$^{-1}$) &   $n_c$   & Signature \\
\midrule
  $2800-3800$  &  $632$  &  OH Band \\ 
  $1730-1800$  &  $41$   &  C=O  \\
  $1660-1670$  &  $6$    &  Peak  \\
  $1530-1580$  &  $29$   &  Trp \\
  $1310-1390$  &  $45$   &  Trp \\ 
\bottomrule
\end{tabular}
\end{table}

In CD and UV absorption spectra, we exclude spectral intervals below 210 nm because their signals are saturated due to the strong absorption of the buffer in this wavenumber region. Consequently, after this spectral truncation, the available data ranges from 210 to 260 nm, corresponding to a total number of spectral components (or spectral intensities) denoted as $n_c = 51$. 

It is worth noting that the spectral truncation applied to CD spectra should address the challenges encountered when attempting to predict the secondary structure of Fib in the presence of CNP using the full spectral range (i.e., 190-260 nm) with the \emph{BeStSel software}~\cite{Micsonai2022, Micsonai2022a}. Specifically, while the software reasonably predicts zero content of $\alpha$-helix structure in bulk Fib at high temperatures, it fails to do so for the protein's spectra in the presence of CNPs (see Fig. S3 in ESI). In this regard, we also found that when regression is performed through Gaussian processes (GPs)~\cite{rasmussen2006, duvenaud2013} on the spectra of Fib with CNPs, the performance deteriorates, indicating an increase in noise in the spectral data. This analysis is reported in Section 4 of ESI.

Finally, we merge the spectra for each system type —bulk Fib, Fib with CNPs, and Fib with SiO$_2$NP— to create new composite spectra that contain significant spectral information, as discussed above. These new spectra are formed by concatenating the UVRR intensities of five spectral intervals (as reported in Table~\ref{tab:table1}), starting from the bottom in ascending order, along with the respective UV intensities, and then appending the CD intensities. As a result, the composite spectra under investigation consist of 856 dimensions.

\subsubsection{Definitions of  Metrics}\label{ML_details}

A central concept in our ML approach is the similarity among different spectra. 
Denoting two spectra by the sets $\left\lbrace (\nu_{1}, I_1), (\nu_{2},I_2),\cdots, (\nu_{N},I_N) \right\rbrace$, $\left\lbrace (\nu_{1},Z_1), (\nu_{2},Z_2),\cdots, (\nu_{N},Z_N) \right\rbrace$,
where $ \nu_{i}$ is the wavenumber and $I_i, Z_i$ the corresponding intensities, their similarity (or equivalently their dissimilarity) will be measured by computing a suitable distance between the vectors: $\mathbf{I} = \left(I_1, I_2, \cdots, I_N \right), \mathbf{Z} = \left(Z_1, Z_2, \cdots, Z_N \right)$, where $N$ corresponds to the number of the spectral observations. 
In Table~\ref{tab:table2}, we report the mathematical formulae of Euclidean and Manhattan metrics, commonly used for assessing the similarity between spectra. 
\begin{table}
    \caption{Formulae of Euclidean and Manhattan metrics}
    \label{tab:table2}
    \centering
    \begin{tabular}{l@{\hskip 1in}l} 
     \hline
    Similarity metric & Formula \\
     \hline\\
     
      Euclidean  & $\left(\sum_{i=1}^{N} \left(I_i - Z_i\right)\right)^{1/2}$  \\\\
      
      Manhattan  & $\sum_{i=1}^{N} |I_i - Z_i|$\\\\
    
      \hline
    \end{tabular}
  
\end{table}

For the sake of completeness, we also report the formula for the cosine similarity $d_{\rm cos}$ between two spectral intensity vectors $\mathbb{I}$ and  $\mathbb{Z}$. It proves beneficial for the computations with t-SNE. 
The cosine similarity reads~\cite{Micsonai2022a}:
\begin{equation}\label{eq:cosine_distance}
d_{\rm cos}\left(\mathbf{I}, \mathbf{Z}\right) = 1 - \frac{\mathbf{I} \cdotp \mathbf{Z}}{\lVert \mathbf{I}   \rVert_2 \lVert  \mathbf{Z} \rVert_2 }  \, ,
\end{equation}
where $\lVert \, \rVert_2$ is the Euclidean norm.

Next, the family of Wasserstein (or Earth Mover’s) distances $W_p$ ($p = 1, \cdots$) arises from the Kantorovich formulation of Monge's mass transport problem~\cite{Monge1781, Kantorovich1948, Kantorovich1960}. The optimal transport distance  $W_p$ is defined between probability distributions for a given metric. 
Thus, given two probability distributions $f\left(\mathbf{x}\right)$ and $g\left(\mathbf{y}\right)$ ($\mathbf{x}, \mathbf{y} \in \mathbb{R}^{n}$), the Wasserstein distance yields the minimum work required to transform one distribution to the other. In the present work, we shall consider the one form of Wasserstein distance $W_1\left(f, g\right)$, that is, $p=1$, which is defined as the minimum of the following expectation value~\cite{prugel2020}:
\begin{equation}\label{eq:wasserstein}
  W_1 \left(f, g \right) =  \min\limits_{\gamma \in \Gamma}  \mathrm{E}_{\gamma}[d\left(\mathbf{x}, \mathbf{y}\right)] \, ,
\end{equation}
where $d$ denotes the Euclidean distance (see Table~\ref{tab:table2}), and $\Gamma$ is the family of of joint probability distributions $\gamma\left(\mathbf{x}, \mathbf{y}\right)$ whose marginal distributions are $f\left(\mathbf{x}\right)$ and $g\left(\mathbf{y}\right)$. This metric is implemented in the Python library SciPy~\cite{2020SciPy-NMeth}.

Applying the Wasserstein distance to spectra requires interpreting them as discrete probability distributions (or, equivalently, discrete probability measures)~\cite{peyre2020}. As a result, the spectral intensities must be nonnegative numbers and normalized to unity. These requirements can be easily met by translating all the multicomponent spectra of a given system and dividing the respective intensities by their total sum. Finally, we note that for computations using the Wasserstein distance, the data do not need to be standardized.

\subsection{ML model}\label{sec12}

The core concept of our ML model is to use a physics-informed metric to identify and quantify protein conformational changes caused by temperature. Because spectral data is high-dimensional, it is essential to visualize emerging patterns and assess their consistency with metric-based analysis. However, handling high-dimensional spectral data effectively remains a significant challenge in current research \cite{10663316}. To achieve this, we employ two state-of-the-art tools: the Wasserstein distance and t-SNE.

These methods are well-suited for high-dimensional data and help overcome the limitations commonly referred to as the \emph{curse of dimensionality}, which include issues like overfitting, sparse distributions, and unreliable distance metrics \cite{james2023, porte2008, Vershynin2018}. Traditional similarity measures in cheminformatics \cite{KHAN201299}, such as Euclidean and Manhattan distances (Table~\ref{tab:table2}), which only consider intensity differences, become ineffective in these conditions, as shown in Section 5 of ESI.

The Wasserstein distance (Eq. \ref{eq:wasserstein}) compares spectra based on their overall shape rather than intensity alone. It calculates the minimal transport cost required to transform one probability distribution into another, making it robust against the saturation of distances in high-dimensional spaces \cite{Thorpe2017}.

For visualization, we use t-SNE, a nonlinear dimensionality reduction technique that maps high-dimensional data into a lower-dimensional space while preserving local relationships. Unlike principal component analysis (PCA), which depends on variance maximization and is sensitive to outliers and variable order \cite{Kim:2021aa}, t-SNE effectively reveals clustering and structure by solving the crowding problem. 
This manifold learning algorithm models similarities between composite spectra in the original high-dimensional space using a Gaussian distribution and maps it into a Student-t distribution in the low-dimensional space \cite{vandermaaten08}. The use of heavy-tailed Student-t distributions in low-dimensional space enables better separation of moderately distant points.
It then minimizes the Kullback-Leibler (KL) divergence between the probability distribution of the original high-dimensional data and the low-dimensional space, which in our case is two, visualizing a corresponding two-dimensional embedding \cite{vandermaaten08}. The KL divergence, which is a non-symmetric measure of similarity between two distributions, is minimized to preserve the local structure of the data. Therefore, if two points are close in high-dimensional space, minimizing the KL divergence ensures they remain close in the low-dimensional space as well.
When initialized with PCA, t-SNE can preserve both global and local structures, resulting in meaningful embeddings \cite{Kobak2019, Kobak2021}.

\section{Results}\label{sec2}
\subsection{Experimental results}

\subsubsection{NP characterisation by Transmission Electron Microscopy and Dynamic Light Scattering without and with Fib}

\begin{figure*}
\centering
\includegraphics[width=\linewidth]{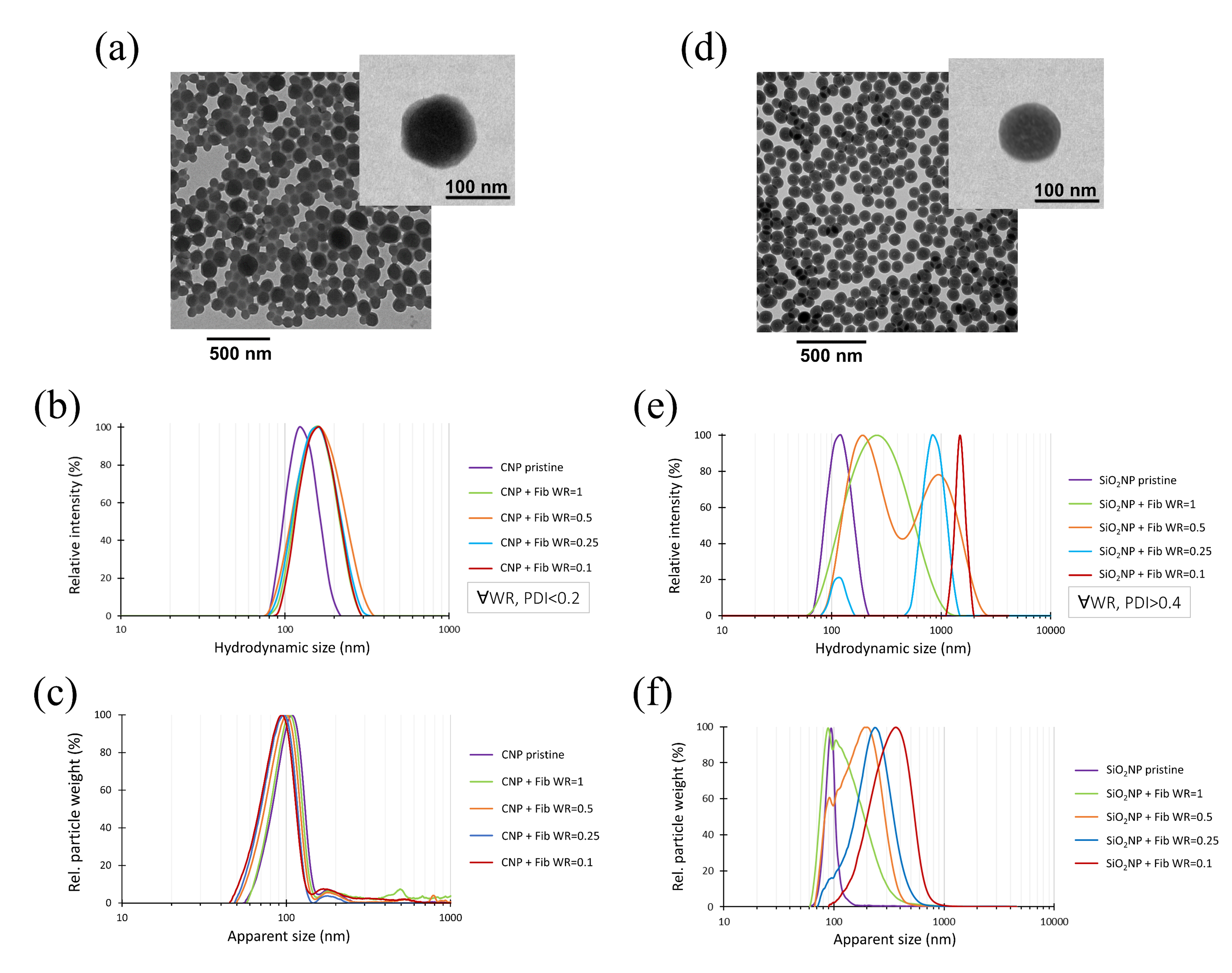}
\caption{Physicochemical characterization of the NPs and the Fib corona.
(a) Transmission Electron Microscopy (TEM) image of CNPs with a nominal diameter of $\sim$120 nm. After image processing, their core diameter was calculated to be (118.3 $\ pm$40.8) nm. For pristine CNPs and CNPs in solution with Fib at WR between 0.1 and 1.0, as indicated in the legend, we measure the DLS distribution (b), consistently finding a polydispersity index (PDI) of less than 0.12, which is consistent with the DCS analysis (c). We repeat the same measurements for SiO$_2$NPs with a nominal diameter of 100 nm: image processing of TEM images estimates their diameter as (95.4 $\pm$ 7.8) nm (d); for pristine SiO$_2$NPs and SiO$_2$NPs in solution with Fib at WR between 0.1 and 1.0, as indicated in the legend, we always observe a PDI of 0.01 by DLS (e), consistent with the DCS analysis (f). Measurements were performed at ambient temperature.}
 \label{fgr:characterization}
\end{figure*}

We follow the experimental protocols developed in our lab to obtain the protein corona formed on top of carbon and silica NPs \cite{soliman2024protocols}. 
We first characterize the pristine NPs by TEM. By using an electron beam on dry samples, TEM provides high-resolution images and characterizes the size populations of NPs (Fig.\ref{fgr:characterization}a and d). To study the samples in colloidal suspension, we use DLS. In DLS, a laser tracks the movement of NPs caused by Brownian motion, enabling measurement of the autocorrelation function and diffusion coefficient, which directly determine the particles' hydrodynamic size distribution and polydispersity index (PDI) in the suspension. We find that both CNPs and SiO$_2$NPs are monodisperse in aqueous solution and remain stable against aggregation (Fig.\ref{fgr:characterization}a and d).

\begin{figure}
\centering
  \includegraphics[width=0.5\linewidth]{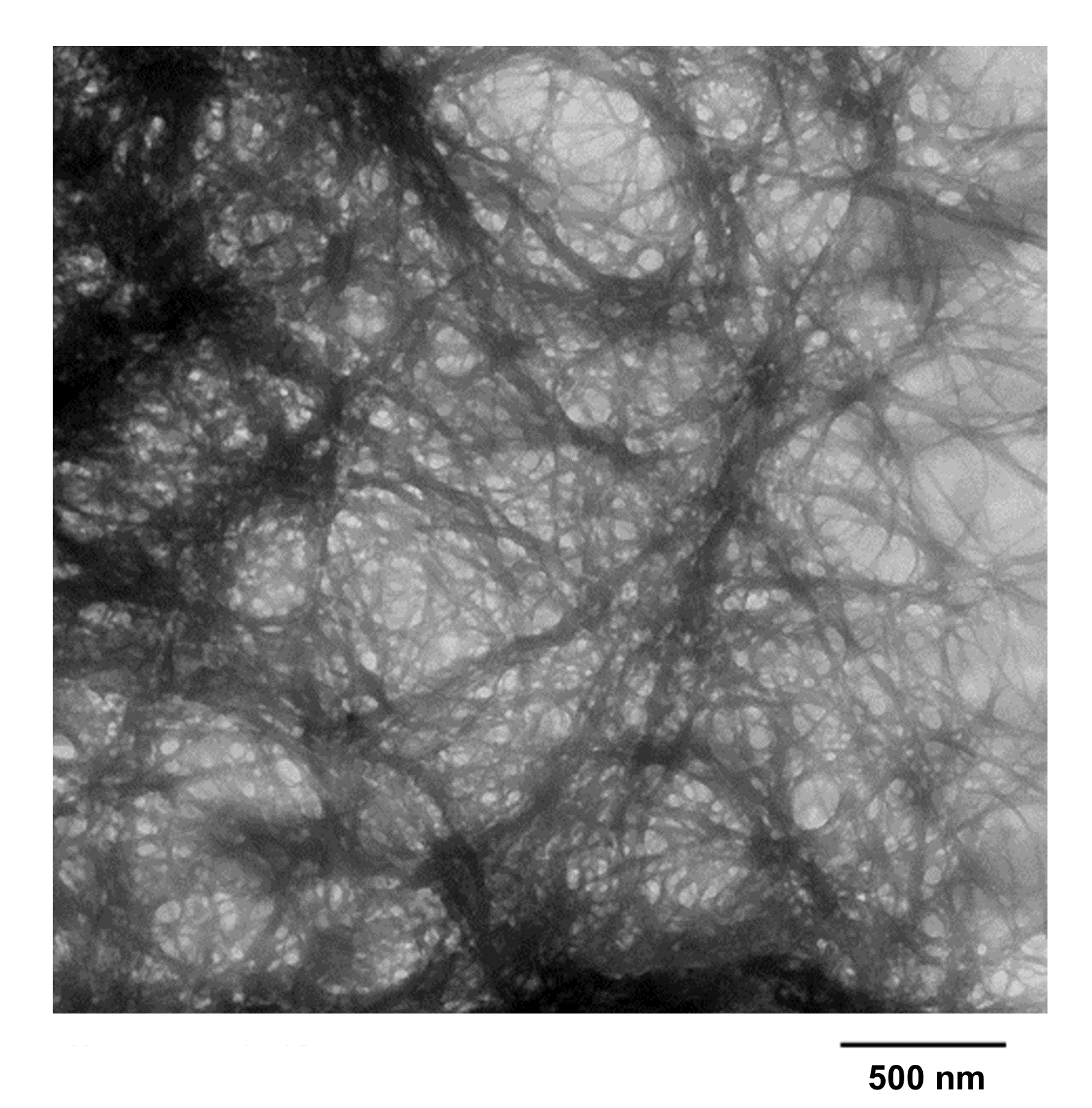}
  \vspace{-1mm}
  \caption{TEM image of Fib aggregates under dried conditions obtained after heating the sample of human Fib at blood-like concentration (2 mg/mL) up to 88$^\circ$C. The shape of the aggregate resembles the typical structure of fibrin fibers.}
 \label{fgr:tem}
\end{figure}

Next, we assess whether the solutions of NPs interacting with Fib are stable at room temperature. In fact, it is well known that adding proteins to a solution can cause NP aggregation \cite{dominguez2016adsorption} or disaggregation \cite{FENOGLIO20111186}, and that aggregation in simple media can occur at much lower protein concentrations than in complex media with many components \cite{doi:10.1021/acsami.2c05362}. 
Our preliminary characterization through TEM shows that Fib at a blood-like concentration agglomerates when the temperature reaches 88$^\circ$C. The aggregates clearly resemble the structure of fibrin fibers \cite{domingues2016thrombin} (Fig.\ref{fgr:tem}). 

We use various techniques to analyze the stability of NP-protein corona complexes against aggregation after incubation at different protein-to-NP weight ratios (WRs), ranging from 0.1 to 1.0. 
By employing DLS and DCS, we do not observe aggregates for CNPs at the tested WRs (Fig.\ref{fgr:characterization} b and c). However, for the same range of WRs, aggregation systematically occurs with SiO$_2$NP (Fig.\ref{fgr:characterization} e and f). 
Hence, the difference in the NPs' chemistry likely impacts protein adsorption and their tendency to aggregate differently. 
For the rest of our study, we set our experimental conditions to WR=1 and compare both CNP (colloidally stable) and SiO$_2$NP (colloidally unstable) with a free Fib solution at various temperatures.

We examine how the hydrodynamic diameter of free Fib and Fib with silica NPs changes with temperature. Both samples show a significant increase in size above 52$^\circ$C (Fig. S1a in ESI). Furthermore, our DLS analysis indicates that, at temperatures larger than 52$^\circ$C, the PDI for both samples rises above 0.4 (Fig. S1b in ESI). These data are consistent with Fib unfolding and possible aggregation above 52$^\circ$C.

In contrast, Fib and CNPs keep a relatively constant hydrodynamic diameter across all tested temperatures (Fig. S1a in ESI), but experience a change in PDI at 52$^\circ$C  (Fig. S1b in ESI). In this case, the two seemingly inconsistent observations do not offer a clear interpretation of the data.

\subsubsection{UV Resonance Raman Spectroscopy}\label{UVRR}

UVRR spectroscopy is a valuable tool for identifying structural changes in organic biomolecules, such as proteins and peptides, without the need for chemical labels \cite{asamoto2019uv}. Compared to spontaneous Raman spectroscopy with visible excitation sources, UVRR offers clear advantages. Proteins' visible Raman profiles often produce complex superpositions of multiple signals, complicating analysis. By tuning the excitation wavelength in the deep UV range (200-300 nm) to achieve selective resonance enhancement, UVRR spectra of proteins can be simplified, making it easier to distinguish signals from specific chromophores or parts of the molecule \cite{oladepo2011elucidating, xu2008hen, jakubek2018ultraviolet}. Using synchrotron radiation (SR) as an excitation source further enhances UVRR experiments by allowing for the precise tuning of the excitation wavelength to achieve optimal conditions \cite{catalini2021hydrogen, catalini2019aqueous, gomez2022amide}. Additionally, UVRR's high sensitivity is crucial for examining proteins in more dilute solutions than those required by visible Raman spectroscopy, all while maintaining high spectral quality. This enables the extraction of information from UVRR spectra of proteins in aqueous solutions, capturing signals from both the polypeptide solute and the water molecules in the hydration shell \cite{punihaole2015uv, catalini2019aqueous}.

\begin{figure*}[t]
 \centering
 \includegraphics[width=\linewidth]{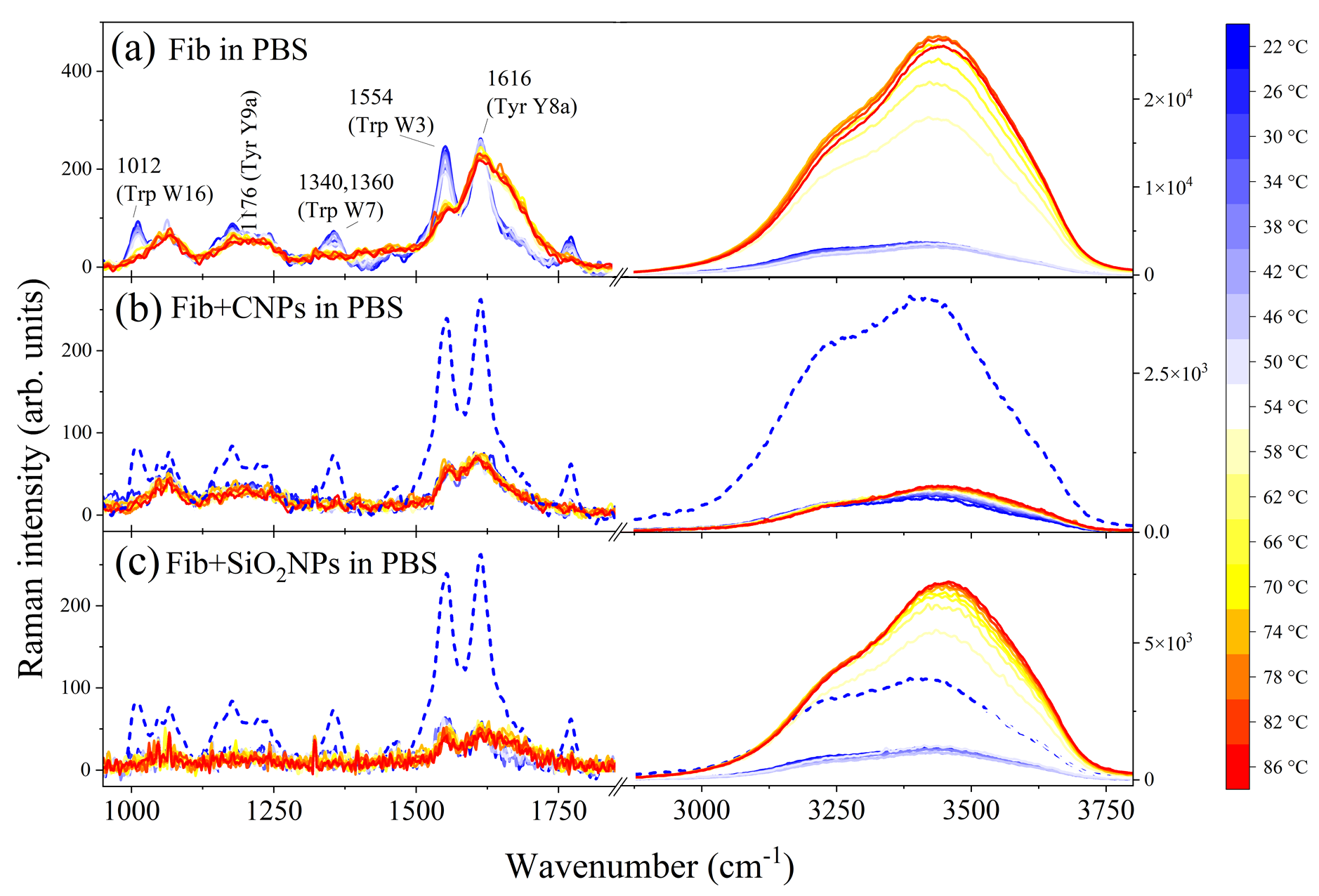}
 \caption{UVRR spectra for free Fib and Fib with NPs. The 226 nm-excited UVRR spectra are collected for Fib at 2 mg/mL in PBS pH 7.4 (a), Fib with CNPs (b), and Fib with SiO$_2$NPs (c) as a function of temperature. These spectra are reported in two ranges: 1000-1800 and 2800-3800 $\mathrm{cm^{-1}}$. The main vibrational features in the Fib spectrum are labeled in panel (a). The temperature range and steps are indicated by the color scale on the right. Dashed lines in panels (b) and (c) show the spectra of Fib alone at the lowest temperature of 22$^\circ$C. 
 }
 \label{fgr:spectra}
\end{figure*}

We collect the UVRR spectra of Fib in PBS buffer, both with and without NPs, using 226 nm as the excitation source (Fig. \ref{fgr:spectra} a-c). This excitation energy falls within the absorption range of the aromatic rings of Tyrosine (Tyr) and Tryptophan (Trp) residues, producing UVRR spectra of Fib that are mainly characterized by the vibrational signals of these aromatic amino acids \cite{oladepo2011elucidating, chi1998uv} (Fig. \ref{fgr:spectra} a). 

In the Raman spectrum of the Fib protein, several key resonance Raman features associated with Trp vibrational modes can be identified, including signals at 1012 $\mathrm{cm^{-1}}$ (Trp W16), attributed to the symmetric benzene/pyrrole out-of-phase breathing; at 1340-1360 $\mathrm{cm^{-1}}$ (Trp W7), which results from the Fermi resonance between N1-C8 stretching in the pyrrole ring and combination bands of out-of-plane bending; and at 1554 $\mathrm{cm^{-1}}$ (Trp W3), related to the C2-C3 stretching mode of the pyrrole ring \cite{cho1994uv}. In the same spectral range of 1000-1800 $\mathrm{cm^{-1}}$, Tyr Raman bands are also observed, specifically the signals at 1176 $\mathrm{cm^{-1}}$ (Tyr Y9a), associated with in-plane C-H bending, and at 1616 $\mathrm{cm^{-1}}$ (Tyr 8a), resulting from the in-plane ring stretching of Tyr \cite{Sweeney:1990aa}. Additionally, a signal around 1770 $\mathrm{cm^{-1}}$ appears in the UVRR spectrum of Fib, likely corresponding to a C=O stretching vibration. 

Since the aromatic rings of Trp and Tyr can participate in hydrophobic, cation-$\mathrm{\pi}$, and hydrogen-bond interactions with nearby residues or surrounding solvent molecules 
\cite{cho1994uv}, the spectral features of the Raman bands associated with these residues offer valuable insights into their local environment. For example, it has been shown that the vibrations of aromatic ring side chains in proteins can act as sensitive markers for changes in the tertiary structure of the proteins during unfolding and aggregation 
\cite{oladepo2012uv, asher1986uv, miura1989tryptophan, schlamadinger2009hydrogen, sanchez2008ultraviolet, C5AN00342C}. 

It is noteworthy that, in the spectral range of 1000-1800 $\mathrm{cm^{-1}}$ (Fig. \ref{fgr:spectra} a), the Raman signal from the water solvent is almost negligible compared to the vibrational signals attributed to the Fib protein. Conversely, the strong spectral distribution characteristic of the OH stretching profile of water completely dominates the high-frequency range between 2800 and 3800 $\mathrm{cm^{-1}}$ in the UVRR spectrum of Fib. 

Remarkable spectral variations can be observed as a function of temperature in both explored wavenumber regions for the protein Fib alone (Fig. \ref{fgr:spectra}a). In particular, the intensity of the Trp Raman signals W16, W7, and W3 significantly decreases starting around 50-60$^\circ$C. For the strong signal at $\sim$1554 $\mathrm{cm^{-1}}$ (Trp W3), the change is even more pronounced, as the peak nearly disappears at higher temperatures. 
Conversely, the signal assigned to the Tyr band at $\sim$1616 $\mathrm{cm^{-1}}$ remains stable with changing temperature, while a new broad spectral feature centered around 1650 $\mathrm{cm^{-1}}$ tends to increase as temperature rises. 
At the same time, we observe a significant increase in the overall intensity of the water's OH stretching band when the temperature surpasses 50-60$^\circ$C. 

The spectra for Fib with CNPs (Fig. \ref{fgr:spectra} b) clearly show that the Raman signal of the Fib protein, when introduced into an NP solution, decreases significantly, and some spectral features almost completely disappear or broaden. For example, this is evident from comparing the spectra of Fib alone and with CNPs in the range of 1500-1800 $\mathrm{cm^{-1}}$. This is partly due to increased UV absorption at 220-230 nm in the solution with CNPs compared to the pristine protein. In fact, self-absorption reduces the observed Raman intensities of Fib, which is clearly seen in the total intensity of the water OH stretching band (Fig. \ref{fgr:spectra} b).  
More importantly, we observe that increasing the temperature does not cause any significant change in the Raman spectra of Fib with CNPs across both explored wavenumber ranges, unlike the case of pristine protein. This finding suggests that the presence of CNPs in solution helps reduce the impact of temperature on Fib's structure. 

\begin{figure}
\centering
  \includegraphics[width=\linewidth]{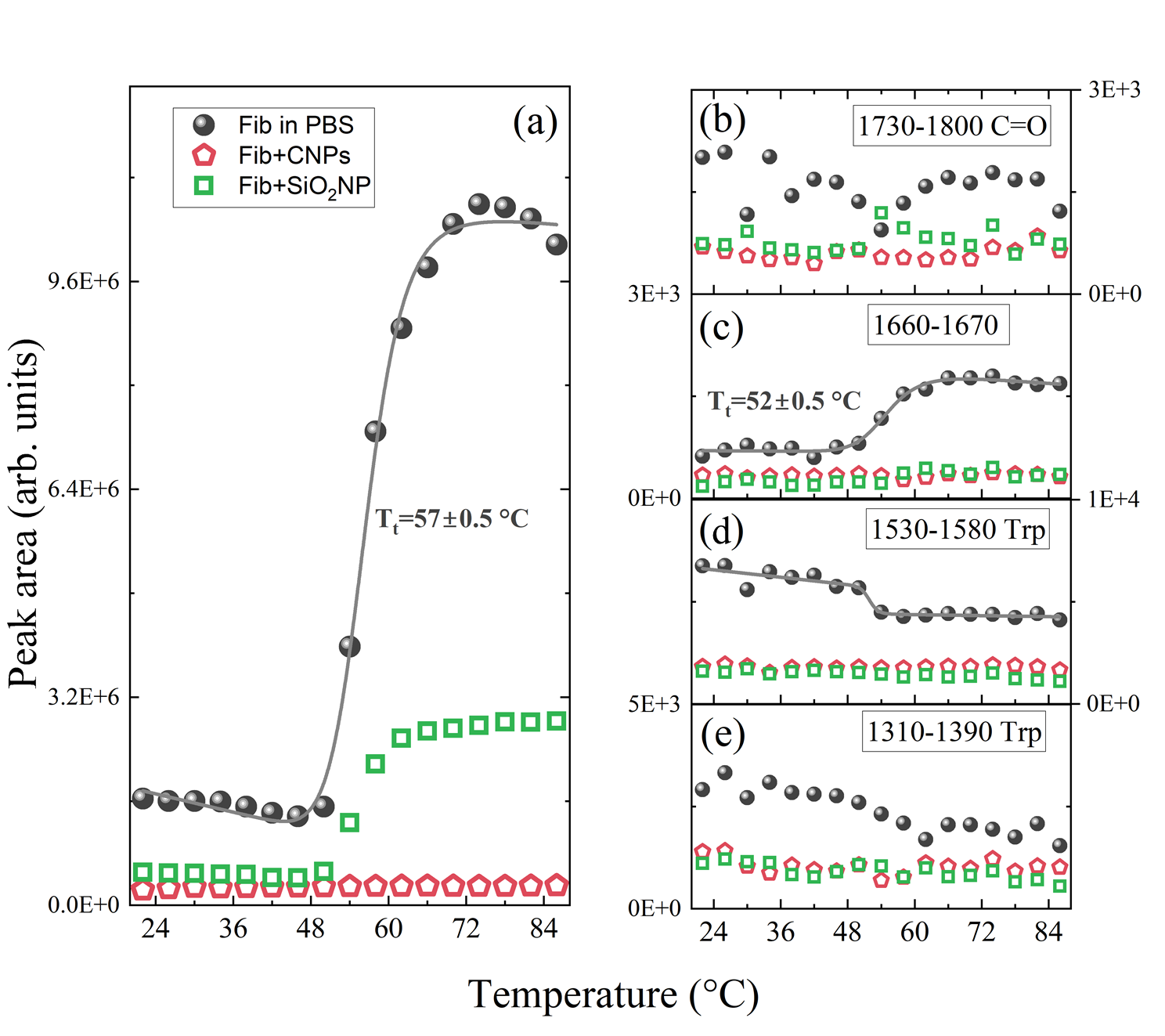}
\caption{Integrated area of some characteristic bands in the UVRR spectra in Fig. \ref{fgr:spectra}. (a) OH band area for Fib in PBS (black circles) and with CNPs (red pentagons), as well as SiO$_2$NPs (green squares). (b)-(e) Areas of the peaks in the frequency intervals reported in Table \ref{tab:table1} for Fib in PBS and with CNPs and SiO$_2$NPs. In each panel, the frequency range is indicated as a label, and the symbols are the same as in panel (a). Where indicated, the experimental data have been fitted with a sigmoidal curve (full black lines).}
  \label{fgr:areas}
\end{figure}

For the spectra of Fib with SiO$_2$NPs, we observe a temperature dependence of the OH stretching band of water that resembles that of free Fib (Fig. \ref{fgr:spectra} c). A more quantitative insight can be gained by analyzing the temperature dependence of the intensities of some characteristic Raman signals (Fig. \ref{fgr:areas}). 

The integrated area of the OH stretching band for Fib alone (Fig. \ref{fgr:areas} a) shows a clear sigmoidal trend with temperature, indicating a transition that occurs at approximately 57$^\circ$C, likely linked to structural changes in the Fib protein. The sharp increase in this Raman signal above 50$^\circ$C likely marks the unfolding and aggregation of the protein caused by the rising temperature. The sigmoidal pattern suggests that protein denaturation begins at approximately 50$^\circ$C and progresses rapidly until 70$^\circ$C, where the OH band area stabilizes. As a result, aggregates may form within this temperature range. 

When CNPs are added to Fib's solution, the OH profile's intensity significantly decreases while maintaining its shape (Fig. \ref{fgr:spectra} b). This may indicate that water cannot penetrate the protein corona, likely due to protein unfolding on the NP surface. 

Additionally, the corresponding OH-band integrated area shows little to no dependence on $T$ (Fig. \ref{fgr:areas} a). This finding suggests that the Fib corona is so strongly associated with particles that even with a significant increase in the solution's energy due to heating, the proteins remain bound to the CNP. Furthermore, the analysis provides no evidence of aggregation.

When SiO$_2$NPs are in solution with Fib, we observe that the OH-band integrated area falls between the two previous cases, with a $T$-dependence that is weaker than that of free Fib (Fig. \ref{fgr:areas} a). Additionally, the increase in area occurs at a temperature close to that of free Fib. This suggests that Fib absorbed onto SiO$_2$NPs undergoes unfolding and aggregation similar to free Fib. However, the analysis does not allow us to quantify the extent of unfolding and aggregation compared to the bulk case.

We also examine how the integrated area of certain characteristic Raman signals in the fingerprint region 1000-1800$\mathrm{cm^{-1}}$, assigned to Fib 
(Fig. \ref{fgr:areas} c-e), depends on the temperature. Specifically, the unfolding and aggregation of proteins are expected to alter the local packing of side chains and the solvent exposure of residues, thereby influencing the environment around Trp amino acids and the resonance Raman cross-section of the related bands \cite{chi1998uv, takeuchi2011uv, ahmed2005uv}. Therefore, by comparing their relative intensity for Fib with and without NPs, we can qualitatively evaluate the effect of NPs on the $T$-dependence of the bands 
(Fig. \ref{fgr:areas} c-e).

Interestingly, for bulk Fib, we observe an increase in the spectral contribution between 1660-1670 $\mathrm{cm^{-1}}$ starting at about 55$^\circ$C (Fig. \ref{fgr:areas} c) and a decrease in the intensity of the Trp W3 band (Fig. \ref{fgr:areas} d) around 52$^\circ$C, while other peak areas show unclear features. These two changes could indicate two different events, such as unfolding and aggregation, or a single event where both occur at the same temperature, as debated in the literature. The experimental uncertainty prevents us from definitively resolving this debate, although there are theoretical reasons to support the first interpretation \cite{Bianco:2020aa}. 

When the solution includes the NPs, none of the band intensities show a consistent trend, making clear interpretation difficult (Fig. \ref{fgr:areas} b-e). Overall, these spectral data highlight the challenge of interpreting the experimental results and emphasize the need for a systematic method for their analysis.

\subsubsection{Circular Dichroism and UV Absorption Spectroscopy}\label{CD_UV}

CD is a common technique in biochemistry for analyzing the secondary structures of biomolecules. By examining the ellipticity $\mathrm{\theta_\lambda}$, which is proportional to the differential absorption of circularly polarized light, researchers can identify how the chiral structures of proteins, such as alpha helices and beta barrels, change under different conditions. Additionally, UV absorbance spectra provide a complementary tool that offers important information about protein stability, unfolding, and aggregation.

\begin{figure}
\centering
  \includegraphics[width=\linewidth]{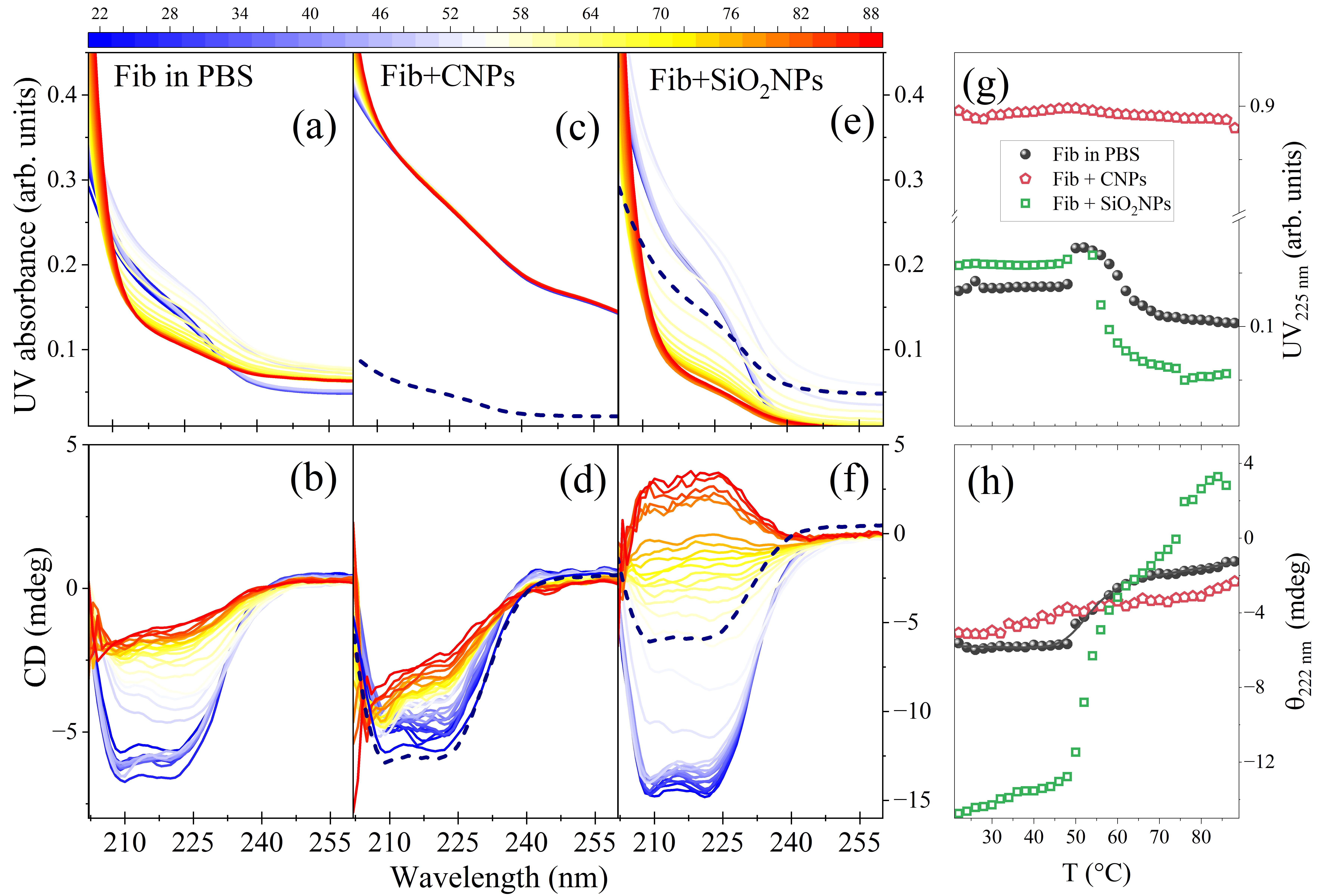}
  \caption{Temperature-dependent UV absorption and CD spectra collected for Fib 0.25 mg/mL under various conditions. In all panels, the temperature ranges between 22 and 88$^\circ$C.
Upper panels a), c), and e) are the UV absorption spectra for Fib in PBS pH 7.4, 
Fib with CNPs, and Fib with SiO$_2$NPs, respectively.
Lower panels b), d), and f) are the corresponding CD spectra.
Curves with different colors correspond to different temperatures as indicated by the upper color bar.
Dashed lines in panels c)-f) represent the spectra of panels a) and b) at the lowest temperature, provided here as a reference. 
Panels g) and h) display the UV absorbance at 225 nm, and the ellipticity $\mathrm{\theta\lambda}$, measured in millidegrees (mdeg), at 222 nm, respectively, for the three cases: free Fib in PBS (black circles), Fib with CNPs (red pentagons), and Fib with SiO$_2$NPs (green squares).}
  \label{fgr:cduv}
\end{figure}

By combining CD/UV absorption spectroscopy with UVRR experiments, we aim to provide a comprehensive view of the structural changes in the Fib protein's secondary and tertiary structures while overcoming the intrinsic limitations of each technique. Therefore, we collect UV and CD spectra as a function of temperature for the Fib protein, both alone and in the presence of NPs (Fig. \ref{fgr:cduv} a-f).

We observe that at approximately 52$^\circ$C in both the CD/UV absorbance spectra of free Fib and Fib with SiO$_2$NPs, there is a noticeable change in the experimental profiles. Specifically, they show i) a decrease in UV absorption (Fig. \ref{fgr:cduv} a and e), and ii) a change in Fib's secondary structure (Fig. \ref{fgr:cduv} b and f of the CD spectra). In particular, the flattening of the CD spectra indicates a loss of ellipticity, which is sensitive to the protein's secondary structure, and the decrease in intensity at 208 nm and 222 nm suggests a loss of $\alpha$-helical regions. These changes correspond to an increase in the water band intensity in the Raman spectra for these two cases (Fig. \ref{fgr:spectra} a and c).

Instead, the UV absorption of Fib with CNPs shows no $T$-dependence (Fig. \ref{fgr:cduv} c), while the CD spectra, which are noisy due to CNP's absorption, exhibit a $T$-dependence that is weaker than those observed for free Fib and Fib with SiO$_2$NPs (Fig. \ref{fgr:cduv} d). These data are consistent with the results from UVRR spectra, which show a very weak dependence of the spectra on temperature (Fig. \ref{fgr:spectra} b).

These trends can be highlighted by selecting signals at a specific wavelength and analyzing them as a function of temperature. For UV absorption, we choose the signal at 225 nm (Fig. \ref{fgr:cduv} g), while for CD, we select the ellipticity $\mathrm{\theta\lambda}$ at 222 nm (Fig. \ref{fgr:cduv} h). For both measurements, we notice some similarities between Fib in PBS and Fib with SiO$_2$NPs, while showing little or no $T$-dependence for Fib with CNPs.

Therefore, similar to the UVRR spectra, although we observe differences and similarities among the three cases, no quantitative results can be systematically extracted from these data, as further discussed in Section 3 (Fig. S3) of the ESI. To analyze the spectroscopic data quantitatively and assess how far the NP-adsorbed proteins deviate from their native state, we apply the ML method that we have developed to the combined spectra.

\subsection{Machine Learning Automate Workflow for High-dimensional Composite Spectra}\label{ML}

To prepare the acquired spectral data for analysis with our proposed ML workflow, we first perform data curation through baseline corrections, followed by suitable spectral truncations. This involves removing silent regions from the UVRR spectra and noisy regions from the CD and UV absorption spectra (see Section~\ref{Data_Curation} for details). The preprocessed spectra for each system (free Fib, Fib with carbon NPs, Fib with silica NPs) are then combined into a higher-dimensional vector that captures all significant chemical information at each temperature.

\begin{figure}
\centering
  \includegraphics[width=\linewidth]{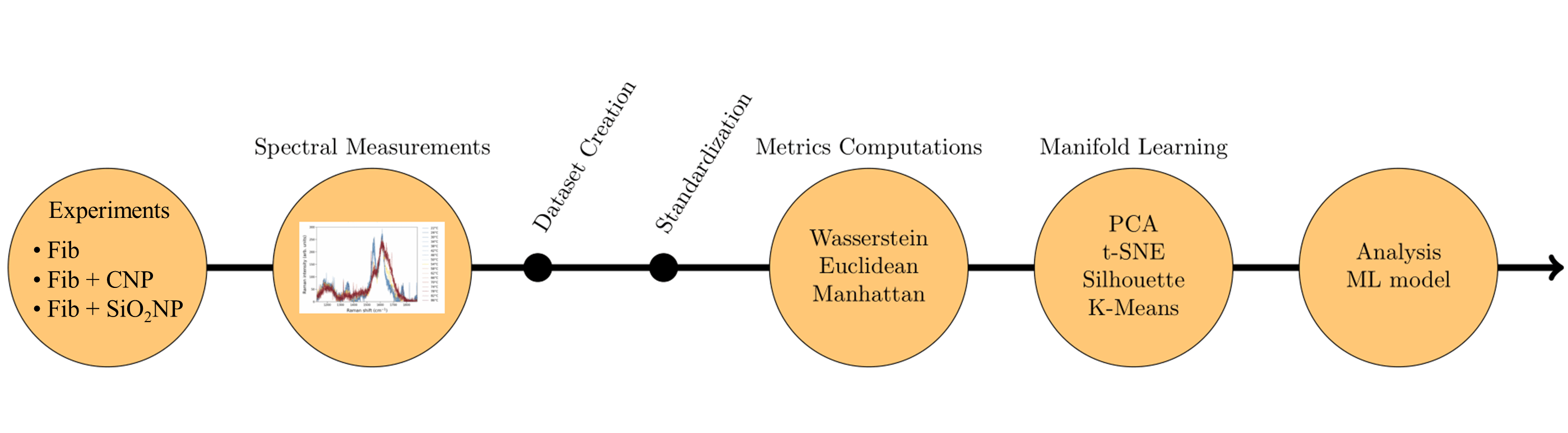}
  \vspace{-1mm}
 \caption{Proposed pipeline including the automated ML workflow. The first two steps (circles) represent the experiment execution and output spectra. In the case considered here, we prepare solutions containing Fib—either alone or combined with CNPs or SiO$_2$NPs—and perform spectral measurements of these systems at various temperatures. Next, we conduct data curation and standardization of the experimental outputs. Then, preprocessed data are analyzed using metrics computation and manifold learning through unsupervised ML tools, enabling both quantitative and qualitative insights analysis.}
  \label{fgr:workflow}
\end{figure}

Next, we implement data standardization, which is usually needed for PCA and t-SNE analysis. Specifically, this step addresses the fact that the data are measured in different units \cite{james2023, Jolliffe2016, Geron2017}.  
Finally, we apply the ML model, which includes (i) metric analysis, (ii) clustering and manifold reduction, and (iii) the subsequent elaboration of the outputs.
%
%
Our entire workflow, from experiments to ML analysis, is shown in Fig.~\ref{fgr:workflow}.
  
\section{Results of the ML model}\label{sec12b}

\begin{figure}
\centering
  \includegraphics[width=\linewidth, height=12cm]{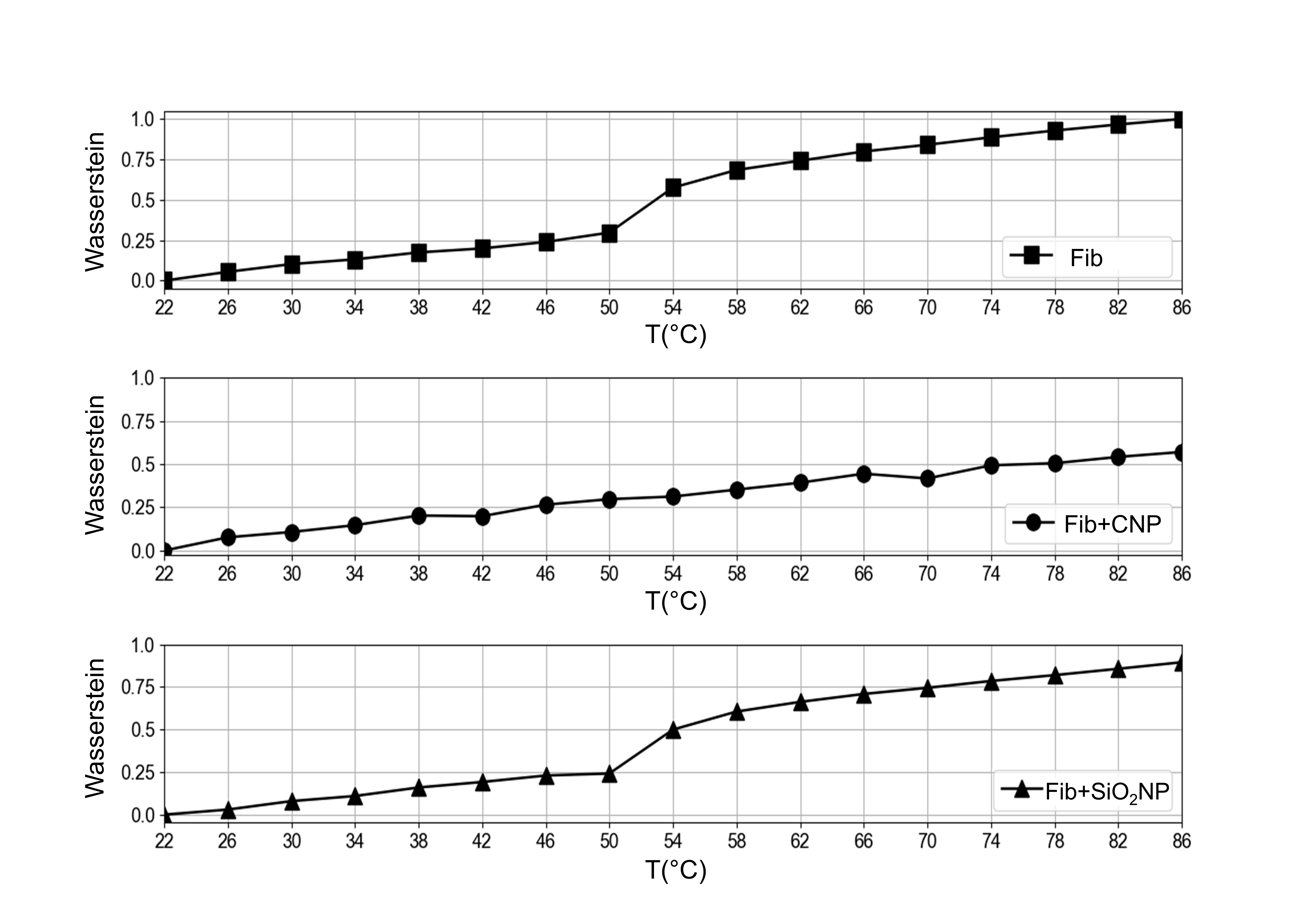}
  \caption{Normalized Wasserstein distance calculated for composite spectra of free Fib and Fib with NPs. Upper panel: free Fib (squares). Middle panel: Fib with CNPs (circles). Lower panel: Fib with SiO$_2$NP (triangles). In each case, the quantity is normalized to the values at $22^\circ$C and $86^\circ$C for the free case, representing Fib in its native (folded) or disordered (unfolded) state. This quantifies the spectral difference (structural disorder) from the reference spectrum of a free Fib protein in its native state at $22^\circ$C.}
  \label{fgr:wasserstein}
\end{figure}

In our metric calculations, we use as reference spectra those recorded at $T_0 = 22^\circ$C, which is the lowest temperature of our measurements. At $T_0$, Fib is in its native state. Consequently, we compute the similarity distance between the composite spectrum at $T_0$ and the spectra at other temperatures $T$, separated by $\Delta T = 4^\circ$C. Accordingly, the Wasserstein distance measures the increase in the protein's structural disorder with temperature. To highlight this change, we normalize the Wasserstein distance using the two values at the extreme temperatures for the free Fib case (Fig.~\ref{fgr:wasserstein}).

We find that the disorder content of the protein increases monotonically with temperature for all systems. Specifically, the normalized disorder content for both free Fib and Fib with silica NPs varies with temperature in a similar way. In both cases, around $(52\pm 2)^\circ$C, the normalized Wasserstein distance undergoes a significant change, while outside this range, the increase is roughly linear.

Comparing this analysis with the original experimental spectra and our knowledge of the temperature range for unfolding the free Fib in solution, we conclude that the characteristic temperature $(52\pm 2)^\circ$C indicates the unfolding transition for both free Fib and the Fib with silica NPs. Therefore, the adsorption of Fib onto SiO$_2$NPs does not affect the Fib unfolding temperature.

However, while the free Fib attains the reference value of one at $T=86^\circ$C, the protein with the silica NP only reaches a value of 0.9 at the same $T$. 
Hence, the silica surface partially limits the unfolding of Fib.

In contrast, the effect of the carbon interface differs significantly. Here, the increase in the normalized Wasserstein distance is roughly linear across the entire temperature range studied, with no sign of a rapid structural change like a folding-unfolding transition. Additionally, the distance at $86^\circ$C only reaches a value of 0.5. Conventionally, a value of 0.5 for a folding-unfolding structural parameter marks the boundary between the two states. Moreover, for the other two cases, the greatest structural change occurs when the Wasserstein parameter is approximately 0.5.
 
Therefore, the impact of the NP surface is significant, causing only a gradual structural change. This analysis helps us better interpret the original spectra, clarifying that the structural disorder induced by increased temperature in this case does not resemble the typical unfolding seen in free proteins. 

These observations about the differences in the Wasserstein distances of Fib proteins interacting with silica NPs and carbon NPs show that the two interfaces influence Fib's structure in fundamentally different ways. Additionally, since we normalize the Wasserstein distance relative to the minimum and maximum disorder in the free Fib case, Fig.~\ref{fgr:wasserstein} provides information about the absolute level of disorder as a function of $T$ in each scenario. Specifically, Fib on silica NPs unfolds about 90\% of the free case, while Fib on carbon NPs is roughly 50\% folded at the same $T$.

It is worth noting that using Euclidean and Manhattan distances would result in a different erroneous scenario, where significant changes occur even at the lowest temperatures across the temperature range. This is discussed in Section 5 of ESI. 

To better understand the total disorder that the Fib acquires as temperature increases, we apply t-SNE to the data in question. 
To compute t-SNE, an input parameter called the perplexity $\mathcal{P}$ is necessary. As discussed in Section 6 of ESI, the perplexity is set to 7, consistent with the small size of the datasets. Other parameters are set to their default values for t-SNE optimization \cite{Kobak2019}.

\begin{figure} 
    \centering
    \begin{subfigure}[t]{0.5\textwidth}
                a)
        \includegraphics[width=\linewidth]{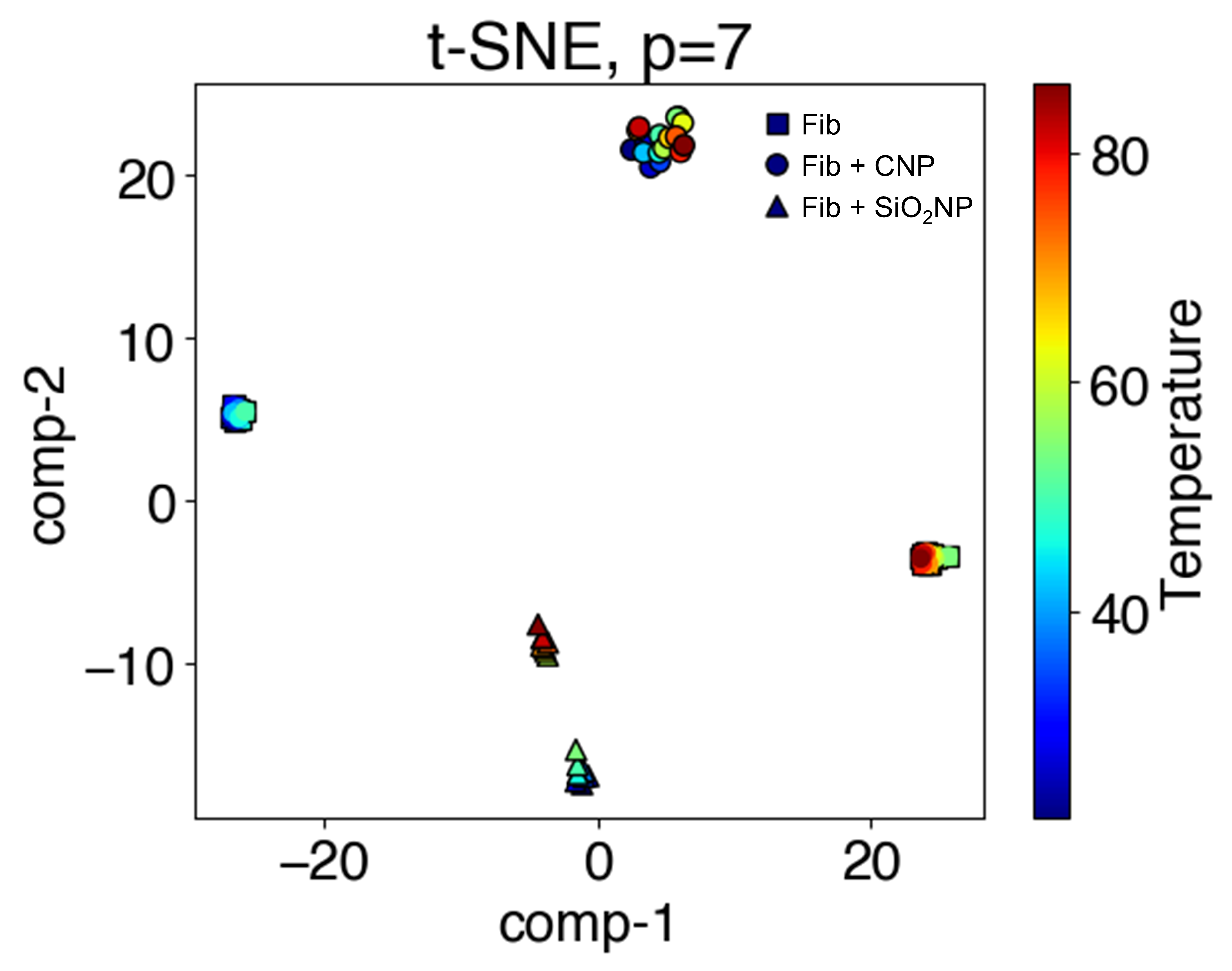} 
    \end{subfigure}
    \hfill
    \begin{subfigure}[t]{0.45\textwidth}
                b)
        \includegraphics[width=\linewidth]{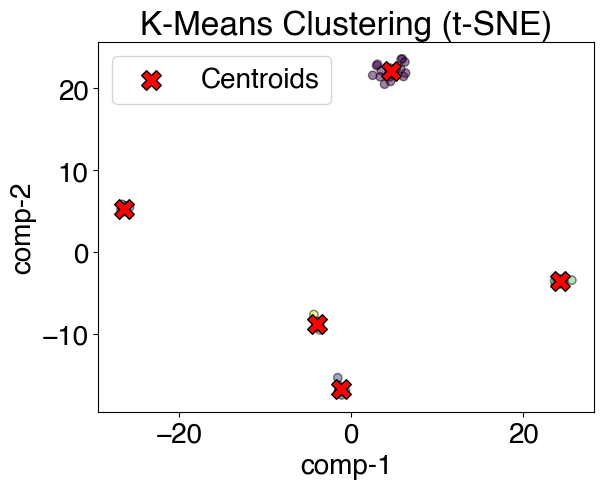} 
    \end{subfigure}
            \caption{t-SNE analysis of composite spectra. a) Two-dimensional embedding from t-SNE with cosine similarity of composite spectra of free Fib (squares), Fib with CNPs (circles), and Fib with SiO$_{2}$NPs (triangles), each corresponding to a specific temperature between $22^\circ$C and $86^\circ$C, at $4^\circ$C intervals. The temperature of each data point is color-coded as shown in the vertical color bar. b) Same as in panel a), with centroids (red cross marks) computed using the K-means clustering algorithm, with an average silhouette score $\approx 0.94$. In both panels, we use $\mathcal{P}=7$, the optimal number of clusters $n_{\rm op} = 5$, and PCA initialization as discussed in Section 6 of ESI.}
\label{fig:tsne}
\end{figure}

To calculate t-SNE, we use cosine similarity (Eq. \ref{eq:cosine_distance}), as it is expected to make t-SNE less susceptible to the high dimensionality of the spectral data. 
We identify five distinct clusters (Fig. \ref{fig:tsne}a). 
All data for free Fib (squares) and Fib with SiO$_{2}$NPs (triangles) are divided into two clusters, each characterized by distinct temperature ranges: $T \leq 50^\circ$C and $T \geq 54^\circ$C. The appearance of two clusters for both free Fib and Fib with SiO$_{2}$NPs indicates a major conformational change around $(52\pm 2)^\circ$C, consistent with our interpretation of the normalized Wasserstein metrics as a measure of protein disorder, which reflects the folding-unfolding transition of Fib as the temperature rises.

The remaining data for Fib with CNPs (circles) are grouped into a single cluster. This indicates that, at all temperatures analyzed, no major structural changes happen, further confirming our rationale that a normalized Wasserstein distance of about 0.5 indicates a protein that is roughly 50\% folded. 

The calculation of centroids using the K-means algorithm, with the number of clusters set to 5, confirms that the data points are well clustered. The clustering pattern has an average silhouette score of approximately 0.94, indicating strong clustering \cite{Geron2017} (Fig.~\ref{fig:tsne} b). 

This analysis demonstrates that, for normalized Wasserstein distance up to 0.5, the change in the Fib structure is limited, as with CNPs. Beyond the 0.5 value, the Wasserstein parameter indicates an unfolded state, as with free Fib and Fib with silica NPs. 

Although not quantitative, the t-SNE analysis qualitatively confirms that the structural change for Fib with SiO$_{2}$NPs is less significant than in the free Fib case, even though the Fib unfolds at the same temperature in both scenarios. In contrast, it highlights the gradual increase in disorder without unfolding for Fib adsorbed on carbon NPs up to $86^\circ$C.

We emphasize that if PCA is used instead of t-SNE, the resulting two-dimensional projection would suggest the presence of five or more clusters. Furthermore, the calculation of the centroids would be less precise, making it more difficult to quantitatively evaluate the absolute amount of structural disorder in the three cases. This is because PCA identifies some data points as outliers, as discussed in Section 7 of the ESI, due to its difficulty in handling data from very different sources.

Finally, we observe that a similar embedding results from computing t-SNE using the Euclidean norm/distance with $\mathcal{P} =7$, as discussed in Section 6 of the ESI. In this case, clustering quality decreases, with the average silhouette score dropping to around $0.91$. Nevertheless, five well-defined clusters still emerge, reinforcing the idea that t-SNE is not significantly impacted by the data's high dimensionality. 

\section{Conclusion}

The study in this article introduces a new method that uses ML to analyze spectral data from multiple experimental techniques. This approach assesses the structural properties of proteins by combining information collected under different conditions. We applied the method to integrated data from UVRR, CD, UV, and DLS measurements. These techniques provide complementary insights into the proteins' secondary and tertiary structures, aggregation tendencies, and overall colloidal stability. 

This study emphasizes the crucial role of NP surface chemistry in affecting the structural stability and aggregation behavior of Fib within the protein corona at different temperatures. Our ML analysis shows that hydrophilic silica NPs reduce Fib’s temperature-induced structural changes by 10\%, while maintaining its characteristic unfolding transition. Above this transition temperature, Fib adsorbed on silica NPs exhibits an aggregation pattern similar to that of the unbound protein, indicating that although the hydrophilic surface moderates unfolding, it does not fully prevent temperature-driven aggregation. In contrast, interaction with hydrophobic carbon NPs keeps Fib in a mildly disordered state across all temperatures studied, without typical unfolding. This interaction notably decreases aggregation, highlighting the stabilizing effect of hydrophobic surfaces against conformational changes and aggregation.

These findings validate theoretical predictions \cite{Fauli:2023aa, march2021protein}, which emphasize how proteins display different unfolding and aggregation behaviors depending on surface interactions. They also highlight the potential to modify protein stability by engineering NP surfaces. Our ML workflow serves as a reliable tool for investigating protein–NP interactions, with significant implications for designing NPs for biomedical applications such as drug delivery and diagnostics. Future research should incorporate additional environmental factors, such as pH and ionic strength, as well as other biomolecules, in more complex biological environments to gain a deeper understanding of protein-NP interactions and enhance their practical utility.

Finally, we emphasize that our ML model offers a flexible framework for dimensionality reduction, which can be further improved to expand its usefulness by incorporating advanced manifold learning techniques. Among these, Uniform Manifold Approximation and Projection (UMAP) has become a strong alternative to t-SNE, offering optimal scalability while preserving both local and global data structures \cite{Yi:2024aa}. Compared to t-SNE, UMAP is at least as computationally efficient and better suited for large datasets, making it especially valuable for high-throughput biological data analysis. Additionally, recent advancements have extended both t-SNE and UMAP for joint visualization of multimodal omics data, broadening their utility in systems biology and precision medicine \cite{Do:2021aa}. These directions promise to enhance the interpretability and analytical power of dimensionality reduction in complex biological systems.

\section*{Acknowledgments}

All authors acknowledge Elettra Sincrotrone Trieste for providing access to its synchrotron radiation facilities and for financial support under the IUS internal project. A.M.-S. and M.P.M. gratefully acknowledge the financial support through the H2020 SUNSHINE grant no. 952924. G.F. and G.M. acknowledge the support by grant PID2021-124297NB-C31 and PID2024-157478NB-C31 funded by  MICIU/AEI/10.13039/501100011033 and European Commission "ERDF A way of making Europe".
G.F. also acknowledges the support from the Spanish Ministry of Universities 2023-2024 Mobility Subprogram within the Talent and its Employability Promotion State Program (PEICTI 2021-2023), the Visitor Program of the Max Planck Institute for the Physics of Complex Systems for supporting a visit started in November 2022, the COST Action CA22143 EuMINe.
G.M. would like to thank Jozsef Kardos for his help in using the BeStSel software. 
All authors thank Oriol Mir\'o for testing the code.

\section*{Authors' contributions}
A.M.-S. performed the experimental measurements and supported the design and implementation of the analysis. G.M. designed and implemented the Machine Learning pipeline/protocol.
F.D. and B.R. provided support for the spectroscopic measurements. B.R. coordinated the activity at Elettra Sincrotrone. I.F. provided the nanoparticles. A.M.-S. and G.M. wrote the first version of the manuscript. A.M.-S., G.M., and B.R. processed the data and prepared the figures. A.M.-S., G.M., B.R., M.M., and G.F. discussed the analysis, the strategies for their elaboration, and the interpretation of the results. G.M., A.M-S., and G.F. designed the computational method and the plan for its implementation. G.M. prepared the support webpage repository for the open-source codes. B.R., M.M., and G.F. prepared revised versions of the manuscript. A.M.-S. wrote the final draft. All the authors discussed and contributed to the final version of the manuscript. M.M. and G.F. acquired funds, designed the study, and coordinated the experimental and theoretical activities.

\section*{Conflict of interest}
There are no conflicts to declare.

\section*{Additional information}

\subsection*{Supplementary information}
The online version contains supplementary material available free of charge.

\subsection*{Availability of data and materials}
The data used in this article can be found at:\\
https://github.com/giancarlofranzese/MASD

\subsection*{Code availability}
The codes developed for this article can be found at:\\
https://github.com/giancarlofranzese/MASD

\bibliography{sn-bibliography}

\end{document}